\documentclass[aps,pre,12pt,onecolumn,epsf,superscriptaddress,graphicx,showpacs]{revtex4}
\usepackage{epsfig,latexsym} 
\usepackage{amsmath,amscd}

\begin{document}

\title{\bf\noindent A comparative study of the energetics of CO on stepped and kinked Cu
surfaces using density functional theory}

\author{Faisal Mehmood$^{\text{1}}$, Abdelkader Kara$^{\text{1}}$, Talat S. Rahman}
\affiliation{116 Cardwell Hall, Department of Physics, 
             Kansas State University, 
             Manhattan, Kansas 66506-2600, USA}
\author{Klaus Peter Bohnen}
\affiliation{Forschungszentrum Karlsruhe, 
             Institut fuer Festkoerperphysik, 
             D-76021 Karlsruhe,
             Germany}
\date{\today}

\begin{abstract}
Our \textit{ab initio} calculations of CO adsorption on several low and high
Miller index surfaces of Cu show that the adsorption energy increases as the
coordination of the adsorption site decreases from 9 to 6, in qualitative
agreement with experimental observations. On each surface the adsorption
energy is also found to decrease with increase in coverage, although the
decrement is not uniform. Calculated vibrational properties show an increase
in the frequency of the metal-C mode with decrease in coordination, whereas
no such effect is found for the frequency of the CO stretch mode.
Examination of the surface electronic structure shows a strong local effect
of CO adsorption on the local density of state of the substrate atoms. We
also provide some energetics of CO diffusion on Cu(111) and Cu(211).
\end{abstract}

\pacs{73.20.---r}

\maketitle

\section{Introduction}

Investigation of CO adsorption on well-defined transition metal surfaces has
been of great academic interest for several decades \cite{blyholder64,bagus83,
davis82,campazona90,wagner87,over01,sung85} 
because of the molecule's obvious relevance to many industrial
processes and as a prototype reactant in studies aiming to provide an
understanding of catalytic reactions \cite{davis82,campazona90,wagner87}.
These experimental and theoretical \cite{over01,sung85} studies
have considered the importance of identifications of 'active sites' based on
examination of adsorption and desorption energies and sticking coefficients
\cite{davis82,campazona90,wagner87}. Since real catalysts consist of small
metal clusters with microfacets of various orientations, the catalyst
surface is generally far from that of one with a low Miller index, rather it
contains defects like steps and kinks which may play specific roles in
determining its reactivity \cite{henry98}. A plausible way to understand
systematically the effect of steps and kinks on chemisorption is to
undertake the examination of CO adsorption on a set of vicinal surfaces, as
has been done in a recent thermal deposition spectroscopy (TDS) study on Cu
surfaces. These experimental studies find a dependence of the CO\ binding
energy on the coordination of the adsorption site, but surprisingly the
effect does not extend to the least undercoordinated sites, namely that of
CO on the kink-sites on Cu(532). Vollmer \textit{et. al.} \cite{vollmer01}
find hardly any difference in the adsorption energy for CO on a step and a
kink site atom implying that coordination alone may not account for the
adsorption energy on this vicinal surface. As proposed by Bagus \textit{et.
al.} \cite{bagus99} in their study of CO adsorption on Cu(100), the
repulsive interaction between the O atom and the \textit{s} states of the
substrate can also play an important role in determining the trends in the
adsorption energy. Similarly in the extensive study of CO adsorption on the
(111) surface of several transition metals, Gajdos \textit{\textit{et. al.}}
\cite{gajdos04} argue that the extent of the shift of the $4\sigma $ and $%
5\sigma $ orbital charge density from the C-O bond to the region below the
carbon (metal surface) may control the adsorption energetics. Of course, an
in-depth understanding of the effect of the local electronic and geometric
structure on CO adsorption may be obtained from the application of
established theoretical methods such as those based on density functional
theory \cite{hohenberg64,kohn65} to the set of Cu surfaces studied
in these experiments. Such calculations may raise some concerns since on the
low Miller index surfaces of Cu they have been found to: a) show preference
for the hollow site \cite{favot01}, while the top site is preferred in the
analysis of experimental data \cite{vollmer01}; b) overestimates the CO
adsorption energy \cite{gajdos04,favot01}. Very recently several
groups \cite{favot01,gajdosss05,mason04} have attempted to
supplement DFT functionals in various ways so as to remedy the two above
shortcomings of DFT for this system with some success. Our intention in this
paper is, however, to carry out a systematic, comparative study of CO
chemisorption on the on-top sites of a set of low and high Miller index
surfaces of Cu to see the trends in adsorption energies as a function of
local coordination and CO coverage. While these calculations have been
motivated by experimental data \cite{vollmer01}, we are also aware of
theoretical work \cite{bagus99,gajdos04,gajdosss05} on CO
adsorption on a few of the surfaces that we are considering here. Naturally,
we will compare our results to all available information that exists to
date. In addition to the trends in CO adsorption energy on Cu surfaces, we
are also interested in the diffusion barrier for CO, since this is a
necessary step in any chemical reaction on the surface. A comparative study
of the diffusion of CO on Cu(111) and Cu(211) is thus presented. Also, given
the interest in reactions at and near steps and kinks on surfaces we have
carried out an examination of CO on several sites near these defects. We are
also including in this work the effect of CO adsorption on the nature of
atomic relaxations of the vicinal surface, to see how the trends in the
relaxation patterns correlate with those in the surface electronic
structure. 

After giving the computational details of our work in next section, we
present a detailed analysis of the calculated CO adsorption energies as a
function of local coordination and their comparison to experiments and other
calculations. This is followed by investigations of the implication of
increasing CO coverage on adsorption energetics. Rest of the paper is
devoted to the characterization of the changes in the surface electronic
structure (local density of states, workfunctions) on CO adsorption. The
vibrational frequencies of CO are also presented on the set of surfaces.

\section{Computational Methods}

\textit{Ab initio} calculations performed in this study are based on the
well-known density functional theory (DFT) \cite{hohenberg64,kohn65}. 
For purposes here, a systematic study of the energetics and the
electronic structure of a set of Cu surfaces of varying geometry was made by
solving Kohn-Sham equations in plane-wave basis set using the Vienna \textit{%
ab initio} simulation package {\small (VASP) } 
\cite{kresse93,kresse96,kresseprb96}. 
The electron-ion interactions for C, O and
Cu are described by ultrasoft pseudopotentials proposed by Vanderbilt 
\cite{vanderbilt90}. 
A plane-wave energy cut-off of 400 eV was used for all
calculations and is found to be sufficient for these systems 
\cite{gajdos04,gajdosprb05}. 
In all calculations reported in this article, the
generalized gradient correction of Perdew and Wang \cite{perdew92} (PW91)
was used since it has been shown to give results for adsorption energetics
and structural parameters which are in better agreement with experiment as
compared to those obtained using the local density approximation (LDA) 
\cite{favot01,fauquet03}. 
The calculated bulk lattice constant for Cu
was found to be 3.6471 \AA\ and the Brillouin zone sampling of the total
energy was based on the technique devised by Monkhorst and Pack 
\cite{monkhorst76} 
for all bulk calculations with a \textit{k}-point mesh of $%
10\times 10\times 10.$

The supercell approach with periodic boundaries is employed to model the
surface systems. To calculate the total energies for several coverages of
CO, we have used surface unit cells consisting of $2-8$ atoms per layer.
While the definition of the coverage for the low Miller index surfaces is
conceptually easy and simply defined as the ratio of number of adsorbate and
substrate atoms in the surface unit cell, for surfaces with steps and kinks
we need also to take into account the specific areas of the unit cells, for
consistency. For vicinal surface we have thus adopted a definition for the
coverage $(\theta ^{\text{vic}})$ which is related to the one on the (111)
surface $(\theta )$ according to $:$ $\theta ^{\text{vic}}=\theta \frac{%
A_{(111)}}{A_{\text{vic}}}$, where $A_{(111)}$ is the area of the surface
unit cell of Cu(111) and $A_{\text{vic}}$ is the area of the vicinal
surface. We have used 2.579 \AA\ as the nearest neighbor distance for
evaluating the area of the surface unit cell.

Cu(100) and Cu(111) surface systems were modeled by a 4 layer (16 atoms)
tetragonal and hexagonal supercell, respectively. These 4 layers are
separated with 11 \AA\ of vacuum. For Cu(100), calculations were performed
for $c(2\times 2)$ overlayer corresponding to 50\% coverage of CO and for a $%
(1\times 1)$ structure corresponding to full coverage even though such a
high coverage is unrealistic. For Cu(111), the $p(2\times 2)$ structure
corresponding to 33\% CO coverage, and those corresponding to coverages of
66\% and 100\% were also studied. For all coverages considered on Cu(100)
and Cu(111), the CO molecule was taken to be adsorbed on the on-top sites,
with the molecule sitting perpendicular to the surface and the C atom
forming a bond with the Cu surface, as also reported in number of
experiments \cite{vollmer01,braun96,graham98}. A $4\times
4\times 1$ Monkhorst-Pack \textit{k}-point mesh was used for Cu(100) and of $%
5\times 5\times 1$ for Cu(111).

The Cu(110), Cu(211) and Cu(221) surfaces were modeled by an orthorhombic
supercell of 8, 17 and 22 layers, respectively, separated with approximately
12 \AA\ vacuum. For Cu(110), a $(2\times 1)$ surface unit cell (16 Cu atoms)
with 1 and 2 CO molecules were chosen to model 50\% and 100\% coverage of
CO, respectively, as shown in the Fig. 1c. A Monkhorst-Pack \textit{k}-point
mesh of $5\times 7\times 1$ was used for this surface. Both Cu(211) and
Cu(221) being regularly stepped surfaces with monoatomic steps and terrace
width of 3 and 4 atoms, needed 34 and 44 atoms, respectively, to model a $%
(2\times 1)$ surface unit cell. For Cu(211), this supercell corresponds to
17.69\% coverage, while for Cu(221) the corresponding coverage is 14.44\%.
Calculations for twice the coverage were also performed by incorporating an
additional CO molecule in the same supercell. The detailed analysis was
performed for the on-top and the bridge site, also for Cu(221), for which
the $(2\times 1)$ supercell corresponds to 14.44\% CO coverage and an
additional atom in the same supercell boosts the coverage to 28.88\%. A
Monkhorst-Pack \textit{k}-point mesh of $5\times 4\times 1$ was used for
(211) and $5\times 3\times 1$ was used for the (221) surface. Finally,
Cu(532) whose surface consists of regularly spaced kinks, was modeled by a
simple monoclinic supercell of five layers, where each layer has 8
non-equivalent atoms. These five layers were separated with 12 \AA\ of
vacuum. A Monkhorst-Pack \textit{k}-point mesh of $3\times 4\times 1$ was
used and CO was adsorbed on top of the Cu atom at the kink site which is the
experimentally proposed preferred site \cite{vollmer01}. Two coverages
(10.76\% and 21.52\%) of CO on Cu(532) were modeled by adsorbing one and two
molecules, respectively, on the kink site (and on a site very next to kinked
site).

Since vicinal surfaces provide local environments with a range of
coordinations and a hierarchy of adsorption sites may exist on them 
\cite{stolbov04}, 
we have carried out calculations for one of the stepped
surfaces, Cu(211), for eight different sites as indicated in Fig. 1d. Since
these results point to a preference for the bridge site, for Cu(221) and
Cu(532) we have calculated adsorption energies for the bridge sites, in
addition to the on-top site.

In order to calculate the total energy of the system for a relaxed
configuration, atoms on all surfaces considered were allowed to move in all
three directions and the structures were relaxed until the forces acting on
each atom were converged to better that 0.02 eV/\AA . The adsorption
energies were calculated by subtracting the total energy of the CO-molecule
and that of the corresponding fully relaxed clean-Cu surface system from the
total energy of CO/Cu surface system $E_{ad}=E_{CO/Cu}-E_{CO}-E_{Cu}$.
Another quantity of interest, the workfunction, was calculated by taking the
difference of the average vacuum potential and Fermi energy for each
surface. Finite difference method was used to obtain vibrational frequencies
of the CO molecule in the gas phase and on the surfaces. CO internal
stretching and CO-metal stretch frequencies were calculated in the direction
perpendicular to the surface. To analyze the nature of the bonding between
CO and the Cu-atoms, local density of states were also obtained. A Gaussian
function of 0.2 eV width was used to smoothen the local density of states
(LDOS).

\section{Results and Discussion}

As already mentioned above, the collection of surfaces studied here provide
variation of adsorption sites with coordination ranging from 6 to 9. The
kink sites with coordination 6 on Cu(532) are particularly interesting
because of their unexpected results from TDS measurements \cite{vollmer01}. 
Also of significance is the fact that the step edges of Cu(211) and
Cu(221) represent the two different microfacets of monoatomic steps on
fcc(111) surfaces. Cu(211) has the (100)-microfacet while Cu(221) has the
(111)-microfacet. It will be interesting to see if such differences in their
geometry affect the energetics of CO on these surfaces. In Table I, we have
compared our calculated CO bond length and the corresponding surface-carbon
distance for these three surfaces. While the C-O bond length remains
insensitive to the local geometry and coordination, C-Cu bond is much
shorter for bridge (b) then ontop (t) site adsorption. As for Cu(532) there
are two non equivalent bridge sites labeled by b$_{\text{KS1}}$ and b$_{%
\text{KS2}}$ in Table I and Fig. 1f. The C-O bond lengths are very close to
the ones calculated by others \cite{braun96,graham98}. When CO is
adsorbed on the bridge site on the three vicinal surfaces, the CO bond
length slightly increases and the carbon-surface distance decreases. Earlier
DFT calculations for CO on Cu(211) also find the same trend \cite{gajdosprb05}. 
The only available date \cite{moler96,andersson79,mcconville86} 
of bond lengths is for the low Miller index surfaces
and is in good agreement with our calculated value of 1.16 \AA\ for the C-O
bond length and 1.85 \AA\ and 1.86 \AA\ C-Cu bond for Cu(111) and Cu(100),
respectively. When compared to vicinal surfaces, we do not see any
difference for C-O bond length and a small variation within 0.02 \AA\ for
C-Cu bonds. A small increase of 0.02 \AA\ was seen in C-Cu bond when CO
coverage was increased from 33\% to 100\% whereas when coverage is doubled
from 50\% in Cu(100) and Cu(110), we see only 0.01 \AA\ increase in C-Cu
bonds and no change was found in C-O bond lengths.

\subsection{Adsorption Energies}

Our calculated CO adsorption energies on various Cu surfaces and their
surface atomic coordination is summarized in Table II. The lowest
coordinated and in turn the most favorable for CO adsorption is the kink
atom on Cu(532) (Fig. 1f), with the highest adsorption energy of 0.98 eV at
a coverage of 10.75\% CO (i.e. one CO molecule/per kink atom). The
calculated adsorption energy for this case is found to be particularly
higher than what has been seen experimentally. Although overestimation of
adsorption energies is typical of DFT based calculations, it is expected
that the qualitative behavior would be similar. For CO adsorption on the
steps of Cu(110), Cu(211) and Cu(221) with atomic coordination 7, our
calculations find $E_{ad}$ to be about $0.86\pm 0.01$ eV, while experimental
values are around 0.6 eV. For adsorption on the kink site we predict an
increase in $E_{ad}$ while experiments \cite{vollmer01} find it to be same
as for the steps. As is clear from Fig. 2 our calculated values scale nicely
with the surface coordination and the variation is larger than that
extracted from experimental data, which is plotted in Fig. 2 for comparison.
The natural question is why this difference between theory and experiment.
We offer here a few rationale. For example, there is the possibility that
site blocking by another CO could lead to smaller measured value on an open
surface like Cu(532). To test the viability of this proposition, we adsorbed
an additional CO molecule on a site next to the kink site (SC1 in Fig. 1f)
which increased the coverage to 21.5\%. As we see in Table II, this leads to
a decrease in the adsorption energy to 0.85 eV which is very close to what
we find for Cu stepped surfaces. The variation of the adsorption energy with
coverage is quite remarkable for all surfaces considered. For Cu(221) $E_{ad}
$ changes from 0.85 eV for 14.5\% coverage to 0.65 for 29\% coverage.
Although DFT studies of CO adsorption on Cu(110) and Cu(211) have already
been reported in the literature \cite{gajdos04,gajdosss05,gajdosprb05,liem04}, 
for consistency we have included them in
Table II. Though we have used the same $(2\times 1)$ cell for Cu(211), this
leads to a different coverage (17.69\%) from that on Cu(221) because of the
difference in the terrace width. The same $(2\times 1)$ cell used for
Cu(110) (with [110] being \textit{x} and [100] being \textit{y}-directions),
as shown in Fig. 1c, results in a 50\% coverage. The CO molecule is adsorbed
on the top site on Cu(110) and Cu(211), as shown in Fig. 1d and 1e. Our
calculated adsorption energies are within 20 meV of each other for all of
cases involving atoms with coordination 7. For Cu(211) our calculated value
is somewhat smaller than in Ref. \cite{gajdos04} which may be associated
with the differences in calculational details. For CO adsorption on Cu(110)
Liem \textit{et. al.} \cite{liem04} obtained an adsorption energy of 0.95
eV for the same coverage and on-top site. The discrepancy with our results
could be due to the usage of fewer layers and smaller \textit{k}-point mesh
and energy cut-off in their calculations. For example, we find the
adsorption energy to change from 0.87 eV to 0.89 eV when we reduce the
number of layers in our supercell from 8 to 6 (note that a three layer slab
was used in Ref. \cite{liem04}). Also, we have found an energy cut-off of
300 eV to be not sufficient to provide converged results for the total
energy for open surfaces such as the (110). A cut-off of at least 400 eV
(value used in all our calculations) is required for accurate determination
of adsorption energies.

For the next surface in this hierarchy Cu(100) whose surface atoms have
coordination 8, we obtain an adsorption energy of 0.77 eV for a 50\% CO
coverage. This result is again higher than the experimental value which
ranges, between $0.5$ and $0.57$ eV \cite{graham98,fohlisch04}. On
the other hand, our result is close agreement with the calculated value of
0.863 eV found for a smaller coverage of 0.25 ML by Gajdos and Hafner \cite
{gajdosss05}, using the same DFT technique. Finally, we turn to Cu(111)
which has a surface atoms with coordination 9 and displays the lowest
adsorption energy of all surfaces discussed here. For this particular
surface, we have used the experimentally studied $p(2\times 2)$ overlayer
which corresponds to 33\% CO coverage. For 0.25 ML coverage an adsorption
energy of 0.74 eV has already been reported \cite{gajdos04}. The small
difference from our result of 0.634 eV may be assigned to the difference in
coverage, and is consistent with a decrease of adsorption energy with an
increase of coverage. The clear trend of increasing adsorption energy with
decreasing local coordination can be seen from the plot in Fig. 2 where
solid triangles are our calculated values and empty triangles are the one
from the experiment \cite{vollmer01}.

As expected we find a hierarchy of adsorption sites on Cu(211). The bridge
site (labeled 2 in Fig. 1d) was found to be slightly preferred over on-top
by $\sim $ 0.06 eV. The sites labeled 3 (fcc-hollow) and 4 (hcp hollow near
step edge), in Fig 1d were the next preferred with adsorption energies of
0.78 and 0.89 eV, respectively. The least favorable site on this surface was
found to be the site labeled 6 which is between two step edges and two
corner atoms and has highest number of bonds with the carbon atom and an
adsorption energy of 0.56 eV. On Site 5, 7 and 8 adsorption energies are
very close to each other ranging from 0.6 - 0.65 eV. The adsorption site
with the highest coordination is the corner atom (CC) on Cu(211) with
effective surface coordination of 10, and labeled as site \# 8 in Fig. 1d
for which we find the adsorption energy to be the lowest of all (0.617 eV),
consistent with the above discussion. Note that we find the bridge site to
be also preferred by about 0.08 eV over the on-top site for CO adsorption on
Cu(221). On the other hand, for Cu(532) which has two non-equivalent bridge
sites referred as b$_{\text{KS1}}$ and b$_{\text{KS2}}$ in Table II and Fig.
1f, we find the adsorption energy to be 0.94 eV which is less than that on
the kink site.

In all cases, adsorption energies drop by $130-250$ meV when the coverage is
doubled for these high Miller index surfaces. Various experiments already
show the strong dependence of adsorption energy on CO coverage on Cu surface
\cite{tracy72,peterson90,truong92}. The electron energy
loss spectroscopy (EELS) experiment of Peterson \textit{et. al.} \cite
{peterson90} shows a 50\% decrease in adsorption energy for 0.3ML increase
in CO coverage for Cu(100).

\subsection{CO diffusion on metal surface}

One of the experimental techniques to determine the adsorption energies is
the temperature programmed desorption (TPD). It is hence desirable to
develop a techniques by which TPD spectra are calculated for a given system.
For the case of low Miller index surfaces, the task may be trivial as only a
limited number of processes are involved. But for the case of real surfaces
with steps and kinks, the situation become more cumbersome. To achieve a
realistic description of TPD spectra from these surfaces, one needs to
calculate not only adsorption energies for sites with different local
environments, but also activation energies associated with different
diffusion paths. As we have reported above, the CO molecules adsorb
preferably near kinks and steps, and it is hence rarely that CO molecules
sit on the down side of the terrace (sites 6 and 8 in Fig. 1d). The relevant
energies are hence those associated with diffusion on Cu(111) and near the
step of a Cu vicinal surface. Here we have in mind systems dominated by the
presence of (111) facets and step edges. The knowledge of these activation
energies along with the different adsorption energies will constitute the
base for kinetic Monte Carlo simulations that will determine the TPD
spectra. In order to reach this goal, the energy landscape for the diffusion
of CO on Cu(111) and Cu(211) surfaces were calculated. Since we were
interested in describing the motion of the CO molecule between the
equilibrium sites, the path on Cu(111) was chosen to be from one top site to
the adjacent one, passing through fcc-hollow, bridge, and hcp hollow sites
(Fig. 3). All layers of the metal surface, and C and O atoms were allowed to
relax fully in all three directions on high symmetry sites. For the
intermediate sites only carbon atom was fixed in one direction and rest of
the system was allowed to relax fully. Three intermediate points between the
top and the bridge site and two points for each bridge - fcc-hollow and
bridge - hcp-hollow were calculated, as shown in Fig.4a. The relative energy
profile of the CO molecule on Cu(111) is shown in Fig. 4a. As in previous
studies, our calculations based on DFT with GGA indicate that CO prefers to
sit on fcc hollow site instead of the top site. However, our calculations
show a barrier of about 200 meV to go from the hollow to the top site and
less than 100 meV to go from the hollow to the bridge site. Fig. 4b and 4c,
show the one dimensional energy landscape for the CO molecule on Cu(211).
Two diffusion paths are chosen in this case: a) top - bridge - top (along
the step edge) and b) top - fcc-hollow (away from the step). Fig. 4b
indicates that diffusion from bridge to top is more favorable than from top
to hollow with diffusion barrier about 60 meV compared about 200 meV barrier
for diffusion from top to hollow site. These results indicate that CO
molecules are more free to roam on the step edge of Cu(211) than away from
it.

\subsection{Interlayer Relaxations}

In Table III, we present a comparison of the interlayer relaxations of the
stepped and kinked Cu surfaces as introduced by the adsorption of the CO
molecule. The results for the corresponding clean surfaces are also
included. The Cu(532) surface has the most dramatic effect of CO adsorption
as $\mathbf{d}_{1,2}$ is found to have a large outward relaxation of $%
+23.5\% $ compared to an inward relaxation of $-17.7\%$ for the clean
surface. The effect is local and as seen from Table III, there is only a
small change in the relaxations for the rest of the layers on CO adsorption.
There is so far no experimental data available for interlayer relaxation on
Cu(532), however, our calculated trend for clean Cu(532) is in agreement
with those calculated with many body potentials \cite{mehmood05}. On the
stepped surface Cu(221), the effect of CO adsorption is also mostly on $%
\mathbf{d}_{1,2}$ in which large inward relaxation of $-16.5\%$ on the clean
surface is overtaken by a small outward relaxation of $+2.9\%$. Calculated
interlayer relaxation on clean Cu(221) matches well with those obtained with
the full-potential linear augmented plane-wave method (FPLAPW) \cite
{desilva04} and many body potentials \cite{sklyadneva98} for this
surface. The interlayer relaxations of clean Cu(211) agrees with those
obtained from low energy electron diffraction measurements and previous
theoretical techniques \cite{gajdosprb05,sondan97,seyller99}. 
Like other two surfaces, Cu(211) shows that the strong contraction of $%
-15.6\%$ between layer 1 and 2 ($\mathbf{d}_{1,2}$) of the clean surface is
overtaken by a small inward relaxation of $+6.4\%$ on CO chemisorption..
This general trend of the change in the top layer relaxation from a large
contraction to expansion upon adsorption of CO is in well accord with the
observation by Ibach and Bruchman \cite{ibach78} who reported a surface
vibrational mode above the bulk band for the vicinal surface Pt(775) (due to
the large contraction) that disappears when CO is adsorbed (reflecting the
change of the large contraction to an expansion).

\subsection{Vibrational Properties and Workfunctions}

Vibrational properties of a `free' CO molecule were calculated by fully
relaxing a single molecule in a large super cell of size, approximately $%
6\times 6\times 22$ \AA . The stretching mode was calculated to be 264.5
meV. This can be compared with the experimentally measured frequency of
isolated CO molecule of 257.8 meV with EELS and of 257.7 meV with infrared
(IR) spectroscopy \cite{hirschmugl90,raval88}. For CO-covered Cu
surfaces, we have calculated the frequencies of two modes: the metal -
molecule ($\nu _{C-Cu}$) stretch and the intra-molecular stretch ($\nu _{C-O}
$). We find a small drop in the vibrational frequency as compared to that of
the gas phase CO due to the bonding on CO with the Cu surface atom. The
frequencies of the CO vibrational modes calculated for all surfaces are
summarized in Table IV. For all cases in which CO is adsorbed on the top
site we see almost the same value of the CO stretching frequency ($\nu _{C-O}
$), as expected. However, there is a small increase in ($\nu _{C-Cu}$), with
a decrease in the coordination of the substrate atom. This is also to be
expected since molecules adsorbed on low coordinated sites will have stiffer
bonds than those on highly coordinated sites. Whereas, for the case of the
bridge site, there is a decrease of approximately $14-16$ meV in CO stretch
frequency, from that for top site, for all cases. In this case understood
due to the molecule strongly bonded with two surface atoms and being not as
free to vibrate as on top site. The drop in $\nu _{C-Cu}$ mode is also $2-4$
meV greater than top site.

In Table V, we have summarized the calculated workfunctions for all Cu
surfaces considered here, with and without the adsorbate along with
available experimental values for clean surfaces of Cu(100), Cu(111) and
Cu(110). Our calculated workfunctions are smaller than the experimental
values of the known surfaces \cite{fall2000,arafune04,mantz71}. 
Our calculated values showed a decrease in workfunctions for
clean surfaces with decrease in coordination of substrate. A similar kind of
behavior can be seen in the experiments for (111), (100) and (110) surface 
\cite{fall2000,arafune04,mantz71}. Our calculations also
showed an increasing trend on workfunction when CO is adsorbed on the
surface. This effect is strongest for Cu(100) surface with workfunction
change of 0.57 eV for all other surfaces this change is less than 0.17 eV.

\subsection{Surface Electronic Structure}

When an electronegative element like C is adsorbed on a metallic surface, it
tends to take charges away from the surface. This would have a strong
dependence upon the coordination of the adsorbate as reported in many
studies \cite{fohlisch2000}. Since we have considered CO adsorption on a
variety of local atomic environments it is important that we examine the
differences in the local electronic structure and the nature of the bonding
of the CO with the substrate. In Fig. 5 - Fig. 7, we have plotted the local
density of \textit{d}-state (LDOS) of Cu atoms and \textit{sp}-states for C
and O atoms to study the \textit{sp-d} hybridization, for three high Miller
index surfaces of interest here. First of all, we see narrowing (Fig. 5 -
Fig. 7) of the \textit{d}-band for the lower coordinated surfaces, as seen
by Tersoff \textit{et. al.} \cite{tersoff81} for their calculations of Ni
and Cu surfaces. The \textit{sp}-state of C and O are split and have almost
filled bonding band. These states make sub-bands with \textit{d}-state of Cu
surface in all case which can be seen to appear at the same low energies as
for CO case. This affect is found to be more localized on flat surfaces but
on stepped surfaces, even though CO is adsorbed on top site we see a small
change in redistribution of LDOS for next layer atom, which can be caused
due to the small inter layer separation i.e. 0.296 \AA\ for (532) as
compared to the biggest 2.1 \AA\ for (111) case. The next layer atom being
so close to CO also experiences some affect of the adsorbate. This could
well describe the different adsorption energies with different coordination.
Along with this electronic charge density distribution and LDOS for these
atoms we can safely say that this difference for flat (high coordinated
sites) and stepped and kinked (low coordinated sites) could serve as the
basis of different adsorption energies.

\section{Conclusions}

In this paper, we have presented the results of a detailed theoretical
investigation of CO adsorption on three low Miller index surfaces, namely,
(111), (100), (110), two stepped surfaces, (211) and (221), and a kinked
(532) surface of Cu. The CO was adsorbed on experimentally observed
preferred sites to explain trends in adsorption energies due to different
geometrical and chemical structure. Our calculated values of adsorption
energies show the increasing trend with decrease in local coordination of
surface atoms with Cu(532) being the most favorable surface for CO
adsorption with lowest coordination. However bridge sites of Cu(211) and
Cu(221) were found to be slightly preferred over top site which was not the
case for Cu(532). We also found a strong dependence of adsorption energy on
coverage of the adsorbate. Our calculations show a large decrease in
adsorption energy with increase in coverage. Our calculations of diffusion
of CO molecule on metal surfaces indicate that it diffuses much more easily
along the step edge as compared to away from the step edge (terrace). A very
small drop in vibrational frequency of free molecule was noted when it was
adsorbed on the metal surface but difference within different surfaces was
negligible due to same on top adsorption site for all surfaces. Workfunction
of adsorbate covered surface was found to increase in all cases compared to
the clean surface. 

\noindent {\bf Acknowledgments}

The authors would like to thank Dr. Sergey Stolbov and Dr. Claude R. Henry
for interesting and very helpful discussions. We also like to acknowledge
financial support from US Department of Energy under grant No.
DE-FGO3-03ER15445 and computational resources provided by National Science
Foundation Cyberinfrastructure and TeraGrid grant No. TG-DMR050018N.

\pagebreak

\begin{table}[tbp] \centering%
\caption{Calculated structural properties of CO adsorbed on the top (t) and
bridge (b) sites on Cu(211), Cu(221) and Cu(532). The two non equivalent bridge
sites on Cu(532) (see Fig. 1f) are labeled as b$_{KS1}$ and b$_{KS1}$. 
Values in square brackets are reported by others [21].
\label{key}} 
\begin{tabular}{|c|c|c|}
\hline
\textbf{Surface} & \multicolumn{2}{|c|}{
\begin{tabular}{cccccc}
\multicolumn{6}{c}{\textbf{Bond Lengths }(\AA )} \\ 
$\mathbf{d}_{C-O}$ &  &  &  &  & $\mathbf{d}_{C-Cu}$%
\end{tabular}
} \\ \hline
\textbf{Cu(211)} & 
\begin{tabular}{c}
1.16 (t)[1.154] \\ 
1.17 (b)[1.168]
\end{tabular}
& 
\begin{tabular}{c}
1.85 (t)[1.84] \\ 
1.51(b)[1.50]
\end{tabular}
\\ \hline
\textbf{Cu(221)} & 
\begin{tabular}{c}
1.16 (t) \\ 
1.17 (b)
\end{tabular}
& 
\begin{tabular}{c}
1.86 (t) \\ 
1.51 (b)
\end{tabular}
\\ \hline
\textbf{Cu(532)} & 
\begin{tabular}{c}
1.16 (t) \\ 
1.17 (b$_{KS1}$) \\ 
1.17 (b$_{KS2}$)
\end{tabular}
& 
\begin{tabular}{c}
1.85 (t) \\ 
1.56 (b$_{KS1}$) \\ 
1.58 (b$_{KS2}$)
\end{tabular}
\\ \hline
\end{tabular}
\end{table}%

\begin{table}[tbp] \centering%
\caption{Variation of the adsorption energies (E$_{ad}$) of CO for two coverages on several Cu surfaces. The 
coverage in the lower entry (in parenthesis) is twice as large as the one for the top 
entry for each surface. In the upper entries the experimental values from 
Ref \cite{vollmer01} are in parenthesis. N$_{NN}$ is the coordination of 
the adsorption site. \label{key}} 
\begin{tabular}{|c|c|c|c|c|c|c|}
\hline
\textbf{Surface} & \textbf{Cu(111)} & \textbf{Cu(100)} & \textbf{Cu(110)} & 
\textbf{Cu(211)} & \textbf{Cu(221)} & \textbf{Cu(532)} \\ \hline
\textbf{N}$_{NN}$ & 9 & 8 & 7 & 7 & 7 & 6 \\ \hline
\textbf{E}$_{ad}$\textbf{\ (eV)} & $
\begin{tabular}{c}
0.634(0.49) \\ 
\\ 
\\ 
0.18{\small (100\%)}
\end{tabular}
$ & $
\begin{tabular}{c}
0.77(0.53) \\ 
\\ 
\\ 
0.18{\small (100\%)}
\end{tabular}
$ & $
\begin{tabular}{c}
0.87(0.56) \\ 
\\ 
\\ 
0.55{\small (100\%)}
\end{tabular}
$ & $
\begin{tabular}{c}
0.86(0.61) \\ 
0.925(b) \\ 
\\ 
0.61{\small (35.4\%)}
\end{tabular}
$ & $
\begin{tabular}{c}
0.85(0.60) \\ 
0.933 (b) \\ 
\\ 
0.65{\small (28.9\%)}
\end{tabular}
$ & $
\begin{tabular}{c}
0.98(0.59) \\ 
0.941(b$_{KS1}$) \\ 
0.949(b$_{KS2}$) \\ 
0.85{\small (21.5\%)}
\end{tabular}
$ \\ \hline
\end{tabular}
\end{table}%

\begin{table}[tbp] \centering%
\caption{Comparison of clean and CO covered $\%$ multilayer
relaxations.\label{key}} 
\begin{tabular}{|c|c|c|c|c|c|c|}
\hline
$\mathbf{d}_{\text{i,i+1}}$ & \textbf{Cu(211)} & \textbf{CO/Cu(211)} & 
\textbf{Cu(221)} & \textbf{CO/Cu(221)} & \textbf{Cu(532)} & \textbf{%
CO/Cu(532)} \\ \hline
$\mathbf{d}_{1,2}$ & $-15.6$ & $+6.4$ & $-16.5$ & $+2.9$ & $-17.7$ & $+23.5$
\\ \hline
$\mathbf{d}_{2,3}$ & $-11.4$ & $-15.7$ & $-1.9$ & $-10.2$ & $-18.9$ & $-18.8$
\\ \hline
$\mathbf{d}_{3,4}$ & $+11.3$ & $+0.2$ & $-15.2$ & $-9.5$ & $-12.7$ & $-16.2$
\\ \hline
$\mathbf{d}_{4,5}$ & $-4.1$ & $+6.9$ & $+18.8$ & $+6.0$ & $-15.0$ & $-12.0$
\\ \hline
$\mathbf{d}_{5,6}$ & $-2.0$ & $-3.9$ & $-2.9$ & $+3.8$ & $-15.0$ & $-16.6$
\\ \hline
$\mathbf{d}_{6,7}$ & $+2.0$ & $-1.8$ & $-6.1$ & $-2.7$ & $-1.4$ & $-3.9$ \\ 
\hline
$\mathbf{d}_{7,8}$ & $-2.1$ & $+1.8$ & $+2.5$ & $+2.0$ & $+1.9$ & $-5.2$ \\ 
\hline
$\mathbf{d}_{8,9}$ & $+0.1$ & $-0.6$ & $+2.3$ & $-4.4$ & $+25.0$ & $+16.4$
\\ \hline
$\mathbf{d}_{9,10}$ &  &  & $-0.9$ & $+2.2$ & $-9.7$ & $+9.6$ \\ \hline
$\mathbf{d}_{10,11}$ &  &  & $+2.0$ & $+1.0$ & $-2.9$ & $-6.6$ \\ \hline
$\mathbf{d}_{11,12}$ &  &  & $+0.8$ & $-1.1$ & $-4.7$ & $-5.4$ \\ \hline
$\mathbf{d}_{12,13}$ &  &  &  & $-0.1$ & $-2.0$ & $-0.6$ \\ \hline
$\mathbf{d}_{13,14}$ &  &  &  &  & $+0.2$ & $-3.3$ \\ \hline
$\mathbf{d}_{14,15}$ &  &  &  &  & $-1.1$ & $-2.9$ \\ \hline
$\mathbf{d}_{15,16}$ &  &  &  &  & $+0.1$ & $+3.7$ \\ \hline
$\mathbf{d}_{17,18}$ &  &  &  &  & $+1.1$ & $-4.2$ \\ \hline
$\mathbf{d}_{18,19}$ &  &  &  &  & $-1.8$ & $+2.3$ \\ \hline
$\mathbf{d}_{19,20}$ &  &  &  &  & $-1.6$ & $-4.8$ \\ \hline
$\mathbf{d}_{20,21}$ &  &  &  &  & $-0.7$ & $+1.1$ \\ \hline
\end{tabular}
\end{table}%

\begin{table}[tbp] \centering%
\caption{Calculated CO frequencies. $\nu_{C-O}$ is for the streching mode of the CO 
molecule and $\nu_{C-Cu}$ is between the surface and the CO-molecule. Here 't' and
'b' represent the top and bridge sites (see Fig. 1). \label{key}} 
\begin{tabular}{|c|c|c|}
\hline
\textbf{Surface} & \multicolumn{2}{|c|}{
\begin{tabular}{ccccc}
\multicolumn{5}{c}{\textbf{Vibrational Frequency (meV)}} \\ 
$\mathbf{\nu }_{C-O}$ &  &  &  & $\mathbf{\nu }_{C-Cu}$%
\end{tabular}
} \\ \hline
\textbf{Cu(211)} & 
\begin{tabular}{c}
251.2 (t) \\ 
236.2 (b)
\end{tabular}
& 
\begin{tabular}{c}
40.3 (t) \\ 
38.4 (b)
\end{tabular}
\\ \hline
\textbf{Cu(221)} & 
\begin{tabular}{c}
251.0 (t) \\ 
235.5 (b)
\end{tabular}
& 
\begin{tabular}{c}
40.1 (t) \\ 
36.7 (b)
\end{tabular}
\\ \hline
\textbf{Cu(532)} & 
\begin{tabular}{c}
252.1 (t) \\ 
235.7 (b$_{KS1}$) \\ 
233.9 (b$_{KS2}$)
\end{tabular}
& 
\begin{tabular}{c}
41.0 (t) \\ 
36.0 (b$_{KS1}$) \\ 
36.4 (b$_{KS2}$)
\end{tabular}
\\ \hline
\end{tabular}
\end{table}%

\begin{table}[tbp] \centering%
\caption{Calculated workfunctions (eV) for clean Cu surface and CO adsorbed
surfaces along with the available experimental values in parenthesis. The experimental
values are taken from \cite{raval88,fall2000,arafune04}.\label{key}}
\begin{tabular}{|c|c|c|c|c|c|c|}
\hline
\textbf{Surface} & \textbf{(111)} & \textbf{(100)} & \textbf{(110)} & 
\textbf{(211)} & \textbf{(221)} & \textbf{(532)} \\ \hline
\textbf{Clean} & $4.89\left( 4.98\right) $ & $4.63\left( 4.65\right) $ & $%
4.49\left( 4.52\right) $ & $4.57$ & $4.58$ & $4.48$ \\ \hline
\textbf{With CO} & $4.93$ & $4.67$ & $4.68$ & $4.72$ & $4.75$ & $4.64$ \\ 
\hline
\end{tabular}
\end{table}\pagebreak%
\begin{figure}
\begin{center}
\includegraphics[width=12cm]{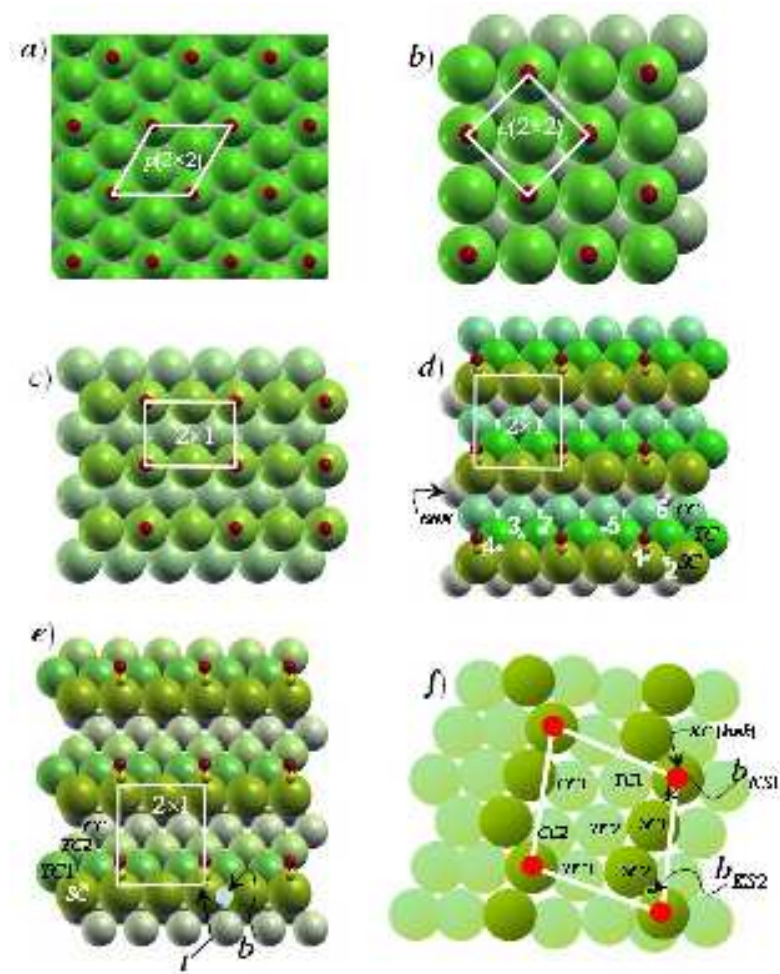}
\caption{
\label{fig: surface}
Top view of a) 111, b)100, C)110, d)211, e) 221 and f) 532 surface. 
Different coverages at different sites are shown. Dark colors represents 
top layer. Note that in 532 surface all eight atoms are non-equalent 
and belongs to eight different layers.
}
\end{center}
\end{figure}
%
\begin{figure}
\begin{center}
\includegraphics[width=10cm]{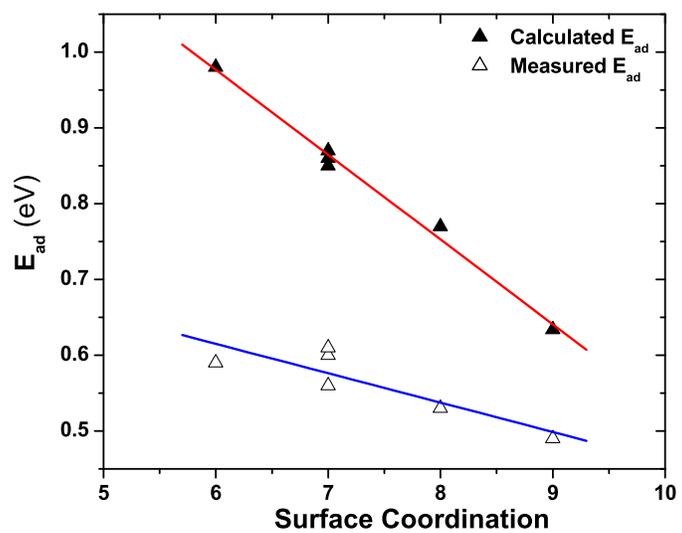}
\caption{
\label{fig: ead vs no bonds}
Adsorption energy versus local surface coordination. 
Solid triangles are calculated values and empty are experimental.
}
\end{center}
\end{figure}
%
\begin{figure}
\begin{center}
\includegraphics[width=14cm]{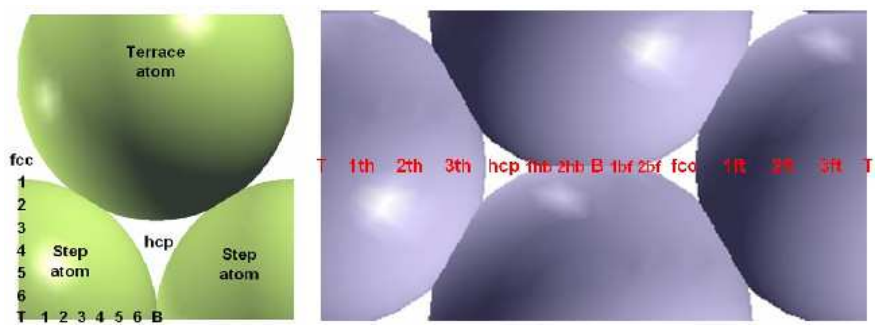}
\caption{
\label{fig: ead vs no bonds}
Chosen diffusion path for CO molecule on a) Cu(111) and b) Cu(211).
}
\end{center}
\end{figure}
%
\begin{figure}
\begin{center}
\includegraphics[width=16cm]{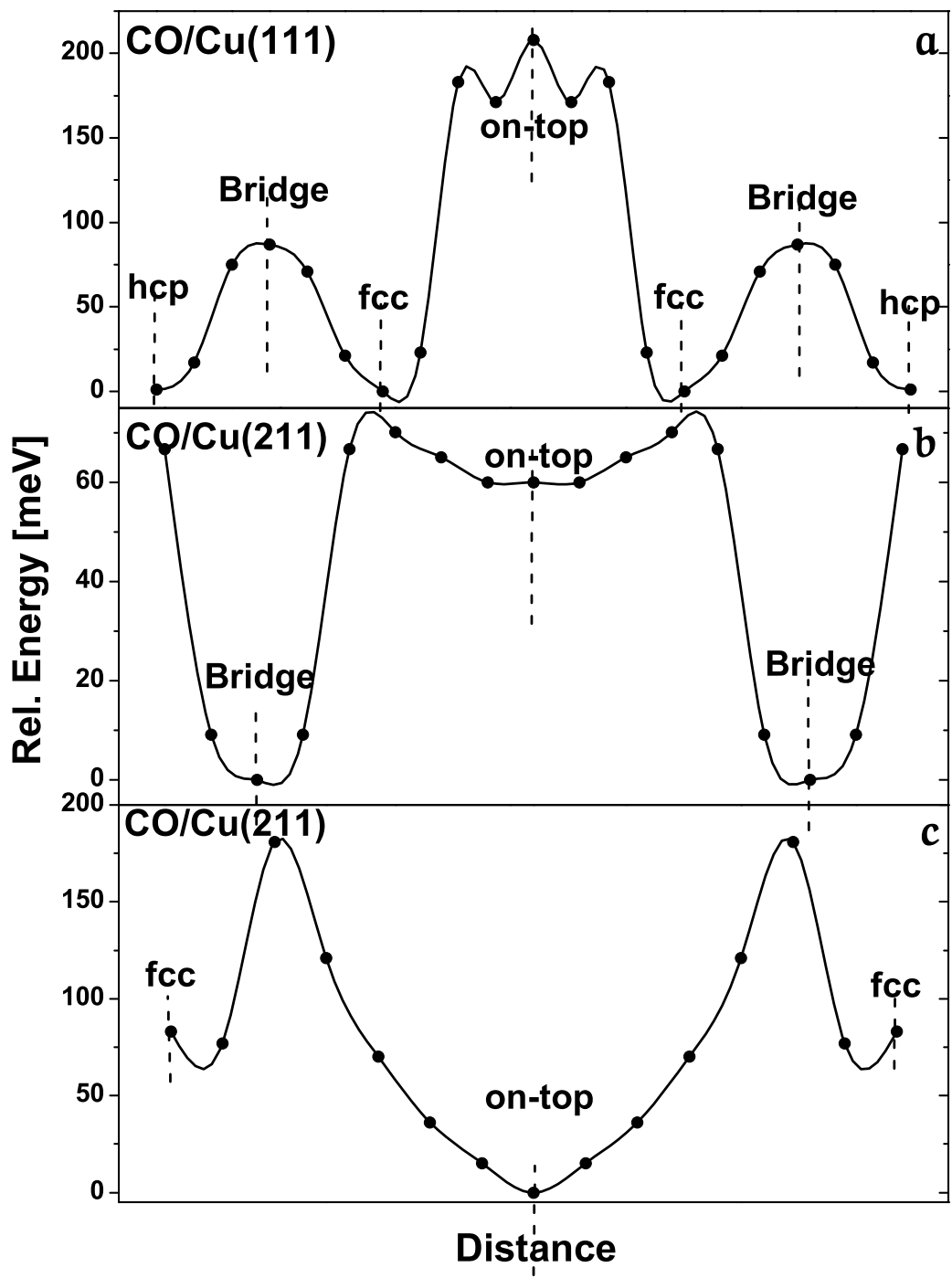}
\caption{
\label{fig: mean and variance of the conductance}
Diffusion of CO molecule on a) Cu(111) and b), c) Cu(211).
}
\end{center}
\end{figure}
%
\begin{figure}
\begin{center}
\includegraphics[width=14cm]{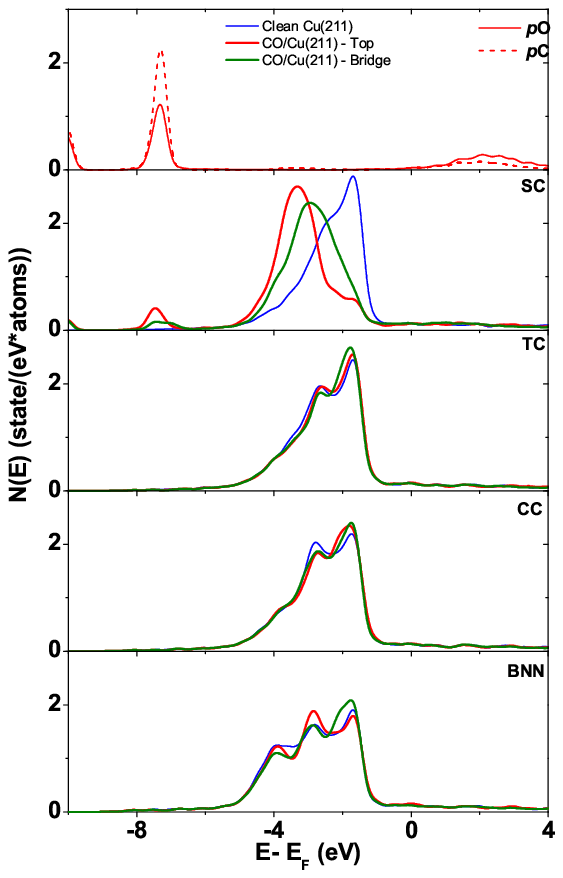}
\caption{
\label{fig: local dos 211}
Local density of state of clean
Cu(211) and with CO on top and bridge sites are shown in Fig. 1d.
}
\end{center}
\end{figure}
%
\begin{figure}
\begin{center}
\includegraphics[width=14cm]{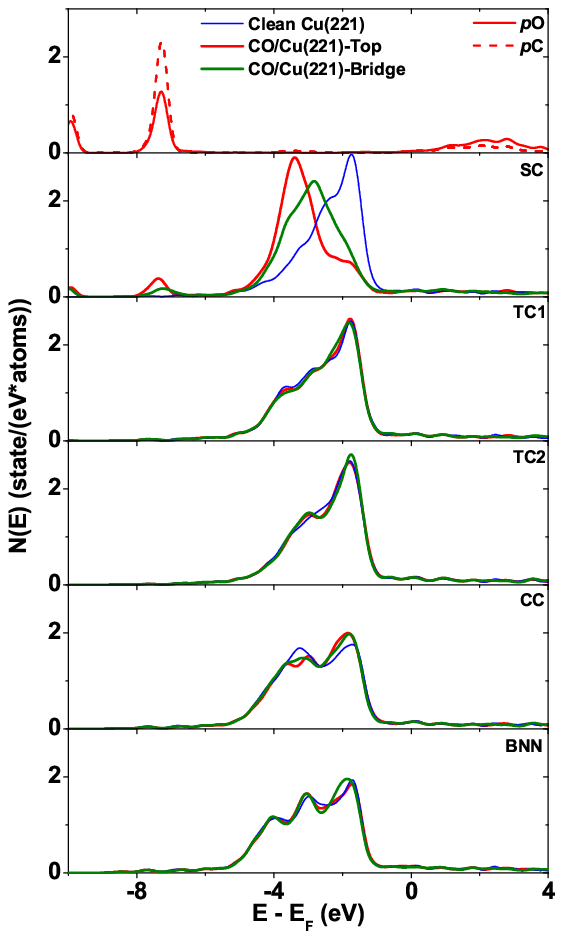}
\caption{
\label{fig: local dos 221}
Local density of state of clean
Cu(221) and with CO on top and bridge sites are shown in Fig. 1e.
}
\end{center}
\end{figure}
%
\begin{figure}
\begin{center}
\includegraphics[width=14cm]{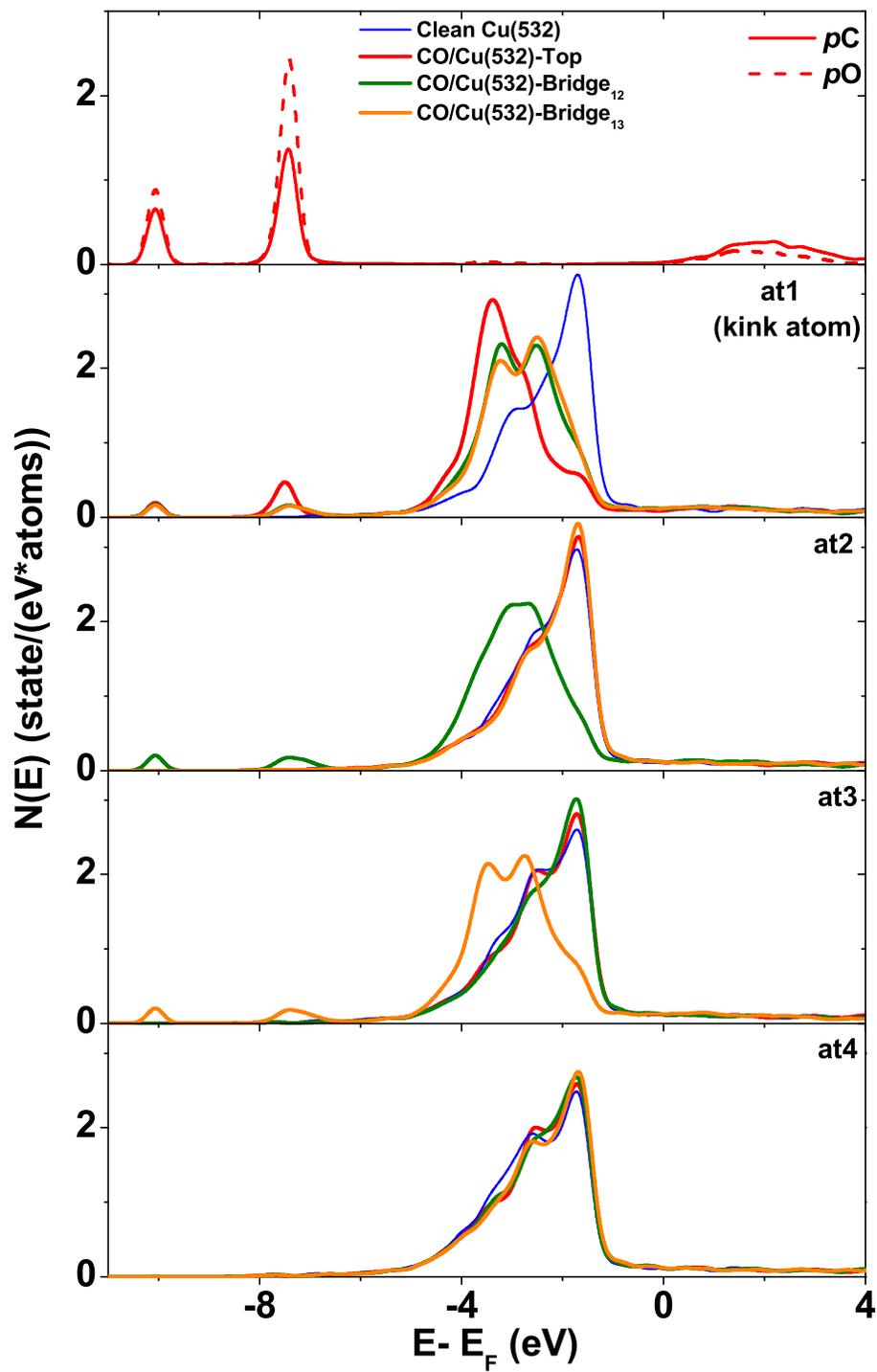}
\caption{
\label{fig: local dos 532}
Local density of state of clean
Cu(532) and with CO on top (kink) and two bridge sites 
are shown in Fig. 1f.
}
\end{center}
\end{figure}
%
\end{document}